\documentclass[journal]{IEEEtran}
\usepackage{booktabs}      
\usepackage{amsmath}       
\usepackage{graphicx}      
\usepackage{float}
\usepackage{orcidlink}

\hypersetup{
    hidelinks
}

\ifCLASSINFOpdf
\else
\fi
\begin{document}
%
\title{Chirped Bragg Gratings as Passive Photonic Reservoirs: An Experimental Proof-of-Concept for Temporal Information Processing}
%
%
%

\author{Isaac Yorke, \orcidlink{0000-0002-1645-6470}\\
 Department of Engineering and Architecture, \\University of Parma, Parco Area delle Scienze 181/A I-43124 Parma, Italy\\
e-mail: isaac.yorke@unipr.it  or  yorkeisaac034@gmail.com}
%
%

\markboth{This manuscript is a preprint and has not been peer reviewed}
{Shell \MakeLowercase{\textit{et al.}}: Bare Demo of IEEEtran.cls for IEEE Journals}
%



\maketitle

\begin{abstract}
Passive photonic structures offer a potential route toward reservoir computing by exploiting intrinsic optical propagation, delay, interference, and memory effects for temporal information processing. This work investigates whether a chirped Bragg grating (CBG) can exhibit fundamental physical properties relevant to a passive photonic reservoir. Rather than presenting the CBG as a complete reservoir computer, the study provides an early proof-of-concept assessment based on experimentally measured device characteristics. Wavelength-dependent group-delay and reflection coefficient measurements obtained from a silicon photonic CBG are incorporated into a distributed delay model in which multiple delayed contributions form the reservoir states. The resulting system is examined with respect to distributed delay, fading memory, temporal mixing, and memory capacity using a linear readout based on the standard memory capacity formulation. The results demonstrate that the measured CBG response provides a distributed temporal memory over the experimentally observed delay range and produces temporal mixing through the superposition of multiple delayed contributions. The memory capacity analysis further indicates that past input symbols can be reconstructed from the distributed delay states, providing quantitative evidence of temporal information retention in the measured CBG response. These results suggest that a passive CBG possesses several physical characteristics relevant to reservoir computing and may therefore provide a simple photonic substrate for reservoir-like temporal processing. The study is intended as an initial proof-of-concept, providing a foundation for future experimental and task specific investigations of passive CBG-based photonic reservoirs. 
\end{abstract}

\begin{IEEEkeywords}
Reservoir computing, Chirped Bragg gratings, Neuromorphic computing, Integrated photonics, Group delay, Photonic computing, Distributed delay, Temporal information processing.
\end{IEEEkeywords}

%
\IEEEpeerreviewmaketitle

\section{Introduction}
%
%
%
%
\IEEEPARstart{P}{hotonic} reservoir computing has attracted considerable interest as a means of exploiting the intrinsic physical dynamics of optical systems for temporal information processing \cite{van2017advances, yorke2026reconfigurable}. In reservoir computing, a high-dimensional dynamical system transforms an input signal into a set of time-dependent states, while a relatively simple readout can be trained to perform a desired computational task \cite{van2017advances, lukovsevivcius2009reservoir, maass2002real}. In photonic implementations, this approach is particularly attractive because optical propagation, interference, dispersion, and nonlinear effects can provide temporal and spatial transformations without requiring every operation to be implemented electronically \cite{van2017advances, abdalla2026photonic}.
An important class of photonic systems for this purpose is represented by passive optical structures in which the physical properties of the device itself provide temporal processing and short-term memory \cite{vandoorne2014experimental}. Such passive reservoirs can exploit optical propagation, interference, attenuation, and delay without requiring active nonlinear elements within the reservoir itself, while nonlinear processing may instead be introduced at the detection or readout stage \cite{ma2021addressing}. Instead, distributed propagation delays, interference, and the resulting temporal response can provide a physical substrate in which information from previous inputs remains accessible to subsequent processing. This motivates the investigation of whether existing passive photonic components can possess reservoir-like properties without being specifically designed as reservoir computers.
Chirped Bragg gratings (CBGs) are particularly interesting in this context because their spatially varying grating period produces wavelength-dependent reflection and propagation characteristics. Different spectral components can therefore experience different reflection positions and group delays, resulting in a distributed temporal response \cite{yorke2024fast, yorke2024analytical}. In addition, the measured reflection spectrum provides wavelength-dependent reflection coefficients that determine the relative strength of the reflected contributions, while the corresponding group delay spectrum assigns a wavelength-dependent delay to these contributions \cite{belai2006group, poladian1997group}. These characteristics suggest that a CBG may provide several of the physical ingredients associated with passive photonic reservoir computing, including distributed delay, temporal transformation, interference between multiple optical contributions, and finite memory of previous inputs.
The present work investigates this possibility as an early proof-of-concept study. The aim is to determine whether a CBG, using experimentally measured device characteristics, exhibits several fundamental properties that are relevant to passive reservoir computing. In particular, the investigation focuses on four aspects: distributed delay, fading memory, temporal mixing, and memory capacity.
The distributed-delay property is examined using experimentally measured wavelength-dependent group delay data obtained from a silicon photonic linearly CBG waveguide. These measurements provide the physical delay distribution used in the subsequent numerical model. The experimentally obtained reflection coefficients are incorporated together with the measured delays to represent the CBG response as a collection of delayed optical contributions. This approach allows the reservoir-like behaviour to be investigated using experimentally derived device characteristics rather than arbitrary delay parameters.
Fading memory is investigated by examining the extent to which information about previous inputs remains accessible as the temporal separation from the input increases. Temporal mixing is subsequently considered by applying a sequence of temporally localized input pulses and examining the resulting superposition of contributions from the distributed delay paths. Finally, the memory capacity is evaluated using the standard linear-reconstruction approach introduced in reservoir computing, in which a trained readout attempts to reconstruct previous input symbols from the reservoir states \cite{dambre2012information, jaeger2001short}.
Together, these analyses provide an initial assessment of whether the measured response of a passive CBG contains the temporal processing and memory characteristics required of a reservoir-like photonic system. The work focuses on identifying and quantifying fundamental physical properties that could form the basis of a passive photonic reservoir. The results are therefore intended as an initial proof-of-concept and as a basis for future investigations involving experimentally implemented reservoir states, nonlinear optical effects, optimized encoding and readout schemes, and task-specific computational demonstrations. The remainder of this paper is organized as follows. Section II presents the device physics and experimental characterization of the chirped Bragg grating. Section III demonstrates the distributed-delay response of the device, followed by the analysis of fading memory in Section IV and temporal mixing in Section V. Section VI evaluates the memory capacity of the CBG using the Jaeger memory capacity framework. Section VII discusses the results, limitations, and potential of the CBG as a passive photonic reservoir, while Section VIII concludes the paper.

\section{device physics and experimental characterization of the chirped Bragg grating}
\begin{figure}[!htbp]
    \centering
    \includegraphics[width=0.45\textwidth]{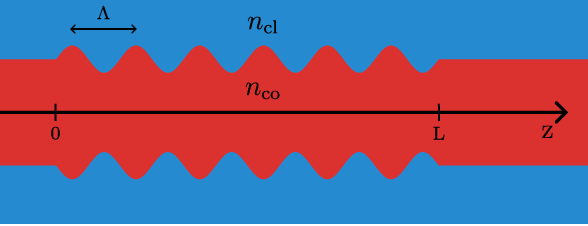}
    \caption{Schematic of a CBG waveguide  \cite{gutt2023integrated}, with the gratings which act as partial reflectors \cite{isaac2025innovative}}
    \label{fig:CBG_Waveguide}
\end{figure}

In this section, an attempt is made to show that the physics already implemented by the CBG performs the same operation as the inner layer of a passive reservoir. Figure ~\ref{fig:CBG_Waveguide} shows the schematic of a CBG waveguide \cite{gutt2023integrated}. In this figure, the waveguide has  cladding and core index of $n_{cl}$ and $n_{co}$ respectively,  the period $\Lambda$ starts at $z=0$ and ends at $z=L$. 

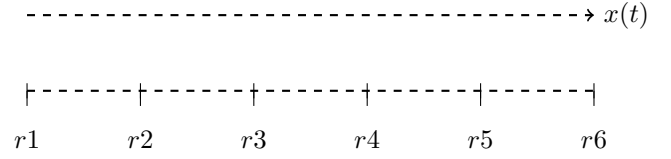
\begin{figure}[htbp]
\centering
\begin{tikzpicture}
    \draw[dashed, ->, thick] (0,0) -- (7.5,0) node[right] {$x(t)$};
    \draw[dashed, thick] (0,-1) -- (7.5,-1);
    
    \foreach \x/\label in {0/r1, 1.5/r2, 3/r3, 4.5/r4, 6/r5, 7.5/r6} {
        \draw (\x,-0.9) -- (\x,-1.1);   
        \draw (\x,-1) -- (\x,-1.2);     
        \node[below] at (\x,-1.4) {$\label$};
    }
\end{tikzpicture}
\caption{Schematic diagram showing the input signal $x(t)$ and the reflections due to the gratings.}
    \label{fig:CBG_reflections}
\end{figure}

From Figure ~\ref{fig:CBG_Waveguide}, it can be seen that a CBG is not one reflector. It is composed of many weak reflectors \cite{isaac2025innovative} and every small section reflects a tiny amount of light. Suppose an input pulse $x(t)$ is injected into the CBG waveguide, the CBG reflects portions of the pulse from many different locations due to the gratings. Figure ~\ref{fig:CBG_reflections} depicts what is happening in the CBG, where grating 1 reflects immediately, grating 2 reflects slightly later, grating 3 reflects even later, and so on.

\begin{figure}[!htbp]
    \centering
    \includegraphics[width=0.45\textwidth]{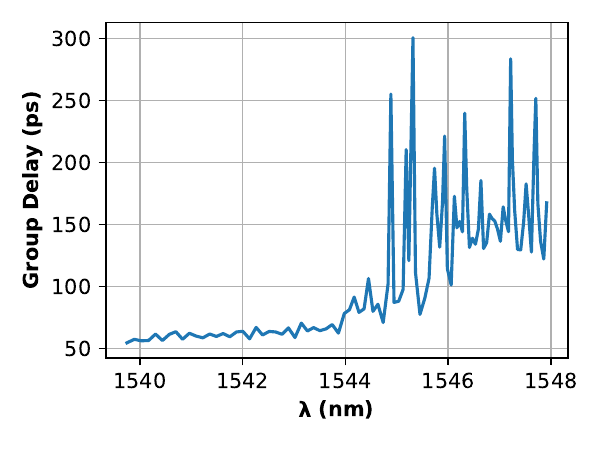}
    \caption{Experimentally measured group delay distribution of the CBG as a function of wavelength}
    \label{fig:group_delay}
\end{figure}

If each of the tiny segments inside the CBG is represented as $dz$, and this tiny segment has reflection coefficient  $r(z)$, the corresponding delay experienced at position $z$ is $\tau (z)$, then the reflected field $dE$ from that tiny segment is:  $dE=r(z)x(t-\tau (z))dz$. The total reflected field is obtained by adding all these tiny contributions along the entire grating. Thus, the total reflected output is given by $y(t) = \int_0^L r(z) \, x(t - \tau(z)) \, dz$. The  discretized version gives: 
\begin{equation}
y(t) = \sum_i h_i \, x(t - \tau_i)
\label{eq:CBG_reflected_output}
\end{equation}. 
\cite{syms1993impulse}, where $h_i$ and $\tau_i$ represent the effective reflection coefficient and delay associated with the $i-th$ section. This represents the simplified distributed delay model motivated by the distributed nature of reflection in a grating. Reservoir computing can be viewed as a dynamical transformation in which an input signal $x(t)$ drives a reservoir, whose internal dynamics generate a set of high-dimensional states $s(t)$. The reservoir is typically characterised by fixed internal connections, represented by a state-transition matrix $W$, while a trainable readout maps the resulting reservoir states to the desired output. Thus, the reservoir does not simply store the input signal; rather, it transforms the temporal input into a rich set of internal states that retain information about the input history and can subsequently be exploited by the readout layer \cite{van2017advances, lukovsevivcius2009reservoir,  dambre2012information}.
Equation \ref{eq:CBG_reflected_output} reveals that every grating position reflects a small amount of light, every reflection has a different delay, every reflection has a different amplitude. So instead of one output, the CBG naturally produces many internal states responses: $s_1=h_1 \, x(t - \tau_1)$, $s_2=h_2 \, x(t - \tau_2)$,..., $s_N=h_N \, x(t - \tau_N)$. Thus;
\begin{equation}
s_i(t)=h_i \, x(t - \tau_i)
\label{eq:reservoir_states}
\end{equation} 
This means the  passive photonic reservoir system can be represented as: $x(t) \rightarrow CBG \ dynamics \rightarrow s(t) \rightarrow y(t)$. The CBG has no explicit matrix. Instead, the physics creates an equivalent operator. Specifically, the collection $\{h_i, \tau_i\}$ plays the role of a high dimensional matrix. So, the complete mapping becomes: $Input \ layer= x(t)$, $Reservoir=\{h_i, \tau_i\}$, $Reservoir \ states=s(t)$ and $	Readout=y(t)$.
Equation ~\ref{eq:CBG_reflected_output} can be written as $y(t) = \sum_i s_i$, where $s_i(t)=h_i \, x(t - \tau_i)$ denotes the contribution of the $i-th$ delayed reflection. The collection of these delayed reflection states forms a distributed set of temporal states that can play a role analogous to virtual nodes in delay-based reservoir computing. In this sense, the CBG provides multiple temporally distinct states directly through its distributed reflection response, without requiring the explicit construction of multiple physical neurons \cite{van2017advances, appeltant2011information}. 
An experimental measurement was carried out at the Technical University of Denmark (DTU) to obtain the reflection spectrum of the CBG. The reflection coefficients $h_i$ were extracted from the measured spectrum, and the group delays $\tau_i$ were subsequently obtained by analyzing the spectrum. Detailed description of how the CBG was fabricated can be found in \cite{gutt2023integrated}. Again, detailed description of how the experimental measurement was carried out, and how the group delay was obtained can be found in \cite{yorke2024fast, isaac2025innovative}. Figure~\ref{fig:group_delay} shows the experimentally measured group delay $\tau_i$ of the CBG as a function of wavelength. The measured group delay spans approximately $50-300$ ps, indicating that the CBG provides a distributed delay response over this range.

\begin{figure}[!htbp]
    \centering
    \includegraphics[width=0.45\textwidth]{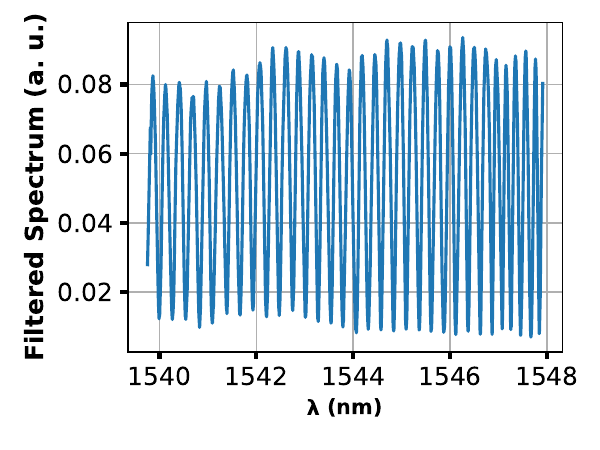}
    \caption{Filtered experimentally measured reflection spectrum of the CBG}
    \label{fig:reflected_spectrum}
\end{figure}

Figure~\ref{fig:reflected_spectrum} shows the filtered experimentally measured reflection spectrum of the CBG. For each wavelength $\lambda$, the corresponding reflection spectrum value was extracted as the reflection coefficient $h_i$, while the group delay analysis at the same wavelength provided the corresponding delay $\tau_i$. Thus, each experimentally determined state is characterised by the pair $\{h_i,\tau_i\}$. Hence, the high-dimensional reservoir state matrix $X$, determined by the pair $\{h_i,\tau_i\}$ is $X=[s_1,s_2,...,s_N]$.

For the numerical demonstrations, unless otherwise stated, the input signal was represented by a sequence of Gaussian pulses with a symbol period of $T_b=50~\mathrm{ps}$ and a pulse width of $5~\mathrm{ps}$. The experimentally measured reflection coefficients $h_i$ and group delays $\tau_i$ were used to parameterise the CBG response.

\section{distributed-delay response}
\begin{figure}[!htbp]
    \centering
    \includegraphics[width=0.45\textwidth]{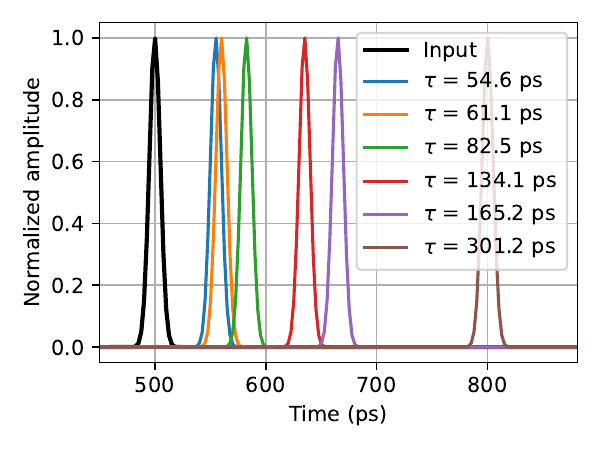}
    \caption{A single input pulse with its delayed replicas due to the CBGs physical delay response. The experimentally measured group delay distribution produces a set of distinct delayed responses to the same input pulse, demonstrating that the CBG provides a distributed-delay substrate for reservoir computing.}
    \label{fig:input_distributed_delays}
\end{figure}

This section demonstrates that the experimentally measured CBG response contains multiple delayed replicas of the same input, with different physical delays $\tau_i$. A single isolated Gaussian pulse centered at $0.5~\mathrm{ns}$ was passed through the CBG. Figure~\ref{fig:input_distributed_delays} shows the plot of the input pulse with the experimentally derived branches $s_i(t)=h_i \, x(t - \tau_i)$  appearing at different arrival times. In this figure, a few of the experimentally measured delay branches spanning from min($\tau$) to max($\tau$) were plotted and it can be seen that the delayed replicas starts from min($\tau \approx 54~\mathrm{ps}$) to max($\tau \approx 300~\mathrm{ps}$), corresponding to the measured minimum and maximum delay distribution in Figure~\ref{fig:group_delay}. This demonstrates that the CBG provides a distributed-delay substrate for reservoir computing.

\begin{figure}[!htbp]
    \centering
    \includegraphics[width=0.45\textwidth]{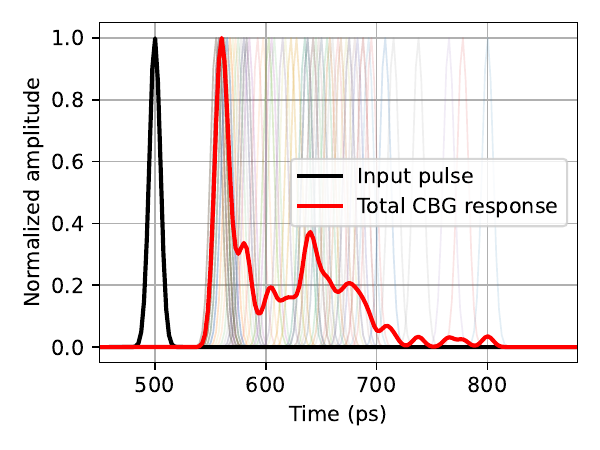}
    \caption{Combined CBG response for a single input pulse and its delay replicas, revealing the physical meaning of distributed-delay in the CBG reservoir output}
    \label{fig:combined_delays}
\end{figure}

Figure~\ref{fig:combined_delays} shows the plot of the combined input and its delayed replicas at the output of the CBG. This reveals the physical meaning of distributed-delay in the CBG reservoir output: $y(t)=\sum_i h_i \, x(t - \tau_i)$, where the experimentally measured $h$'s and $\tau$'s are different.

\section{fading memory }
\begin{figure}[!htbp]
    \centering
    \includegraphics[width=0.45\textwidth]{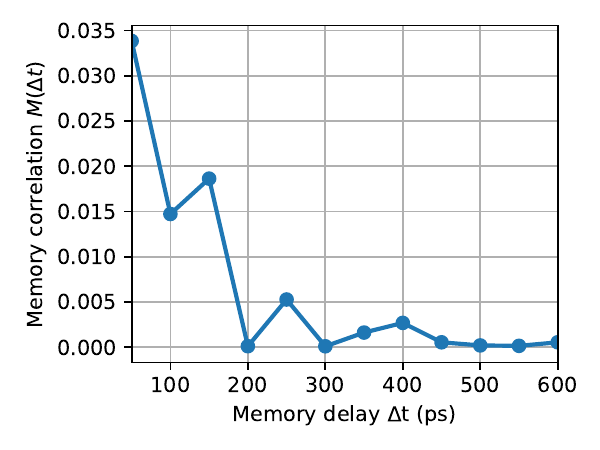}
    \caption{Fading memory behaviour of the system which demonstrates that the correlation between the CBG output and increasingly delayed versions of the input becomes substantially weaker.}
    \label{fig:fading_memory}
\end{figure}

This section demonstrates how strongly does the output correlate with an old input. The correlation-based fading memory strategy was adopted to demonstrate the fading memory behaviour that originates from the measured distributed delays  of the device. The fading memory correlation is therefore calculated as: 
\begin{equation}
M(\Delta t)=corr^{2}[y(t),u(t-\Delta t)]
\label{eq:MC_fading_mem}
\end{equation}
\cite{jaeger2001short}. For the sequence of Gaussian pulses generated, if $k$ represents the arrival index of the discrete symbols, then  $\Delta t=kT_b$ represents the physical delay within the CBG. Thus; for $k=[1,2,3,...]$, in the physical system, $\Delta t=[50,100,150,...]~\mathrm{ps}$. Figure~\ref{fig:fading_memory} shows the fading memory behaviour of the CBG which basically reveals that recent inputs are remembered well, whilst older inputs gradually disappear. In this plot, it can be seen that beyond the maximum experimentally measured delay ($\sim 300~\mathrm{ps}$), the correlation is $\sim 0$. The CBG has a distributed set of delays $\tau_i$  and weighted responses $h_{i} x(t-\tau_i) $, therefore the response contains contributions at different delays. The correlation can consequently have small local maxima and minima and the plot  shows this quite clearly. 

\section{temporal mixing }
\begin{figure}[!htbp]
    \centering
    \includegraphics[width=0.45\textwidth]{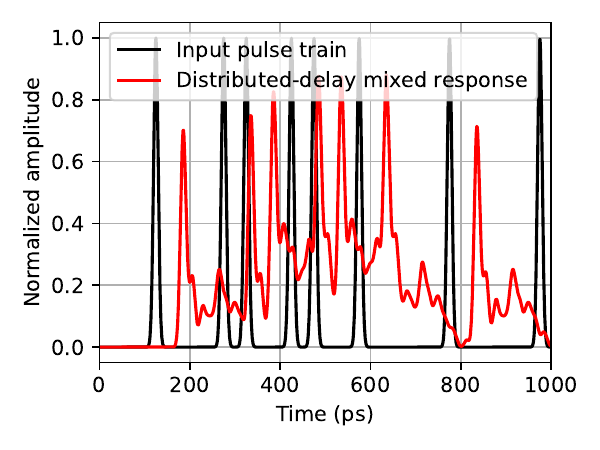}
    \caption{Temporal mixing illustration produced by the CBG's distributed delays. The input is a sequence of Gaussian pulses with $T_b$ and pulse-width as explained. The output is a superposition of many delayed copies.}
    \label{fig:Temporal_mixing}
\end{figure}

Many photonic reservoir papers design a delay network \cite{van2017advances, larger2017high, van2020demonstrating}, the CBG is already exhibiting one. Again, among other requirements, reservoir computing relies on three key properties: memory, high-dimensional representation, and mixing. For memory, the reservoir must remember previous inputs \cite{yorke2026photonic}, and the CBG does this because $x(t-\tau_i)$ contains past versions of the signal. For high-dimensional representation, the reservoir should transform the input into a set of internal states spanning a higher-dimensional state space \cite{duport2016fully}. The CBG does this because one pulse becomes ${s_{1},s_{2},...,s_{N}}$. Thus, one input has become many delayed states. For the mixing operation, the reservoir states should combine information from different temporal instances of the input \cite{rodriguez2019optimal}. The CBG naturally sums $s_{i}(t)$; $y(t)=\sum_i s_i$, which is exactly a mixing operation. For a sequence of input pulses, the output is a superposition of many delayed copies, consequently, the output at any instant contains information from multiple past input times. This is precisely temporal mixing in delay-based photonic systems. Figure~\ref{fig:Temporal_mixing} is a plot that illustrates the qualitative behaviour of the temporal mixing produced by the CBG's distributed delays. In the plot, it can be seen that output response is no longer a sequence of isolated copies of the input. Instead, the different delayed contributions overlap and produce a temporally extended, irregular waveform. 

\section{memory capacity}
\begin{figure}[!htbp]
    \centering
    \includegraphics[width=0.45\textwidth]{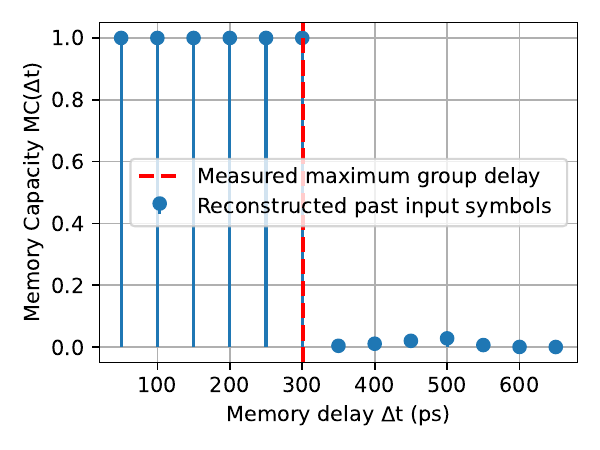}
    \caption{Memory capacity of the system, which shows a near-perfect reconstruction of past input symbols over the experimentally measured delay range.}
    \label{fig:mem_capacity_Jaeger}
\end{figure}

In this section, the memory capacity, introduced in \cite{jaeger2001short}, to measure how accurately the current reservoir state can reconstruct past inputs is adopted. Suppose the input sequence is $u(n)$, the reservoir receives this sequence and produces states $s(n)$. We then ask: Can the readout reconstruct the input from k time steps ago using the reservoir states? The memory capacity is therefore defined as:
\begin{equation}
{MC}_k = {corr}^2\left( \hat{u}(n-k), \, u(n-k) 
\right)
\label{eq:MC_Jaegar}
\end{equation}
where $u(n-k)$ is the reconstructed reservoir and $\hat{u}(n-k)$ is the readout output. The ridge regression readout strategy was employed, the training and testing sequences were 5000 each, and the test sequence was independent of the training sequence. Figure~\ref{fig:mem_capacity_Jaeger} shows the plot of the system's memory capacity as function of physical delay. For $T_{b}=50~\mathrm{ps}$ and a delay distribution of $50 - 300~\mathrm{ps}$, Figure~\ref{fig:mem_capacity_Jaeger} illustrates essentially, a ${MC}_k$ of 1 for the first 6 symbol delays. This suggests that the model employed to describe the CBG is making the system behave as a distributed tapped-delay reservoir, where some reservoir states directly expose past input symbols \cite{ikuta2024optical}. The memory capacity analysis demonstrates near-perfect reconstruction of past input symbols over the experimentally measured delay range and the memory  falls away outside the measured delay window. This behaviour arises from the distributed-delay structure of the CBG. Consequently, the observed memory capacity is interpreted as the linear recoverability of past inputs from the distributed-delay state representation rather than as evidence of nonlinear memory processing. 

Equation~\ref{eq:MC_fading_mem} describes the fading memory analysis whilst Equation~\ref{eq:MC_Jaegar} describes the memory capacity analysis. The differences between the 2 relations is explained in Table~\ref{tab:fading_mem_vs_mem_capacity}.

\begin{table}[htbp]
\centering
\caption{Differences between fading memory and memory capacity of the CBG}
\label{tab:fading_mem_vs_mem_capacity}
\begin{tabular}{p{4cm} p{4cm}}
\toprule
\textbf{Fading Memory } & \textbf{Memory Capacity } \\
\midrule
Uses the aggregate output $y(t)$ & Uses the individual reservoir states $s_{i}(t)$\\
Correlates output with delayed input & Trains a readout for each delayed input \\
Asks how much delayed information is present in the aggregate response & Asks whether delayed information can be reconstructed from the state vector\\
Qualitative memory analysis& Quantitative linear memory capacity measurement \\
\bottomrule
\end{tabular}
\end{table}

\subsection{pulse-width sensitivity analysis}
For the input Gaussian pulse considered in this work,
\[x(t) = A \exp\left[-\frac{(t-t_0)^2}{2\sigma^2}\right]\] 
the width parameter is the Gaussian standard deviation $\sigma$, and therefore the corresponding FWHM is $2\sqrt{2\ln2}\times \sigma$ \cite{klingbeil2015theoretical}. For $T_{b}=50~\mathrm{ps}$, Table~\ref{tab:pulse_width_analysis} shows some selected $\sigma$, the $MC$ obtained in the simulation, and the corresponding FWHM. From the table, it can be observed that narrow pulses ($\sigma=3-5~\mathrm{ps}$) yielded a memory capacity of approximately $6$ symbol delays. Increasing the pulse width progressively increased the calculated memory capacity, reaching approximately $8$ at $\sigma=30~\mathrm{ps}$. This increase is attributed primarily to temporal overlap between neighbouring Gaussian symbols, which introduces correlations into the encoded input sequence and can facilitate delayed-symbol reconstruction. Consequently, a narrow pulse width of $5~\mathrm{ps}$ was adopted for the principal memory capacity characterization to minimize inter-symbol overlap

\begin{table}[htbp]
\centering
\caption{Pulse width sensitivity analysis table}
\label{tab:pulse_width_analysis}
\begin{tabular}{p{1cm} p{1cm} p{1.4cm}}
\toprule
\textbf{$\sigma~\mathrm{ps}$} & \textbf{$MC$} & \textbf{$FWHM~\mathrm{ps}$} \\
\midrule
1 & 5.0017 & 2.36\\
3 & 6.0014 & 7.07\\
5 & 6.0014 & 11.78\\
10 & 6.7904 & 23.55\\
15 & 7.0006 & 35.33\\
20 & 7.0006 & 47.10\\
25 & 7.9651 & 58.88\\
30 & 8.0006 & 70.65\\
\bottomrule
\end{tabular}
\end{table}

\subsection{symbol period sensitivity analysis}
For a fixed Gaussian width of $5~\mathrm{ps}$, a symbol-period ($T_{b}$) sweep was performed and Table~\ref{tab:symbol_period_analysis} shows the $T_{b}$ with their corresponding $MC$. From the table, it can be seen that, from $T_{b}=30-100~\mathrm{ps}$, the results follow $MC\approx300/T_{b}$. So, with an effective memory timescale ($\tau_{memory}$) of $300~\mathrm{ps}$, $MC\approx \frac{\tau_{memory}}{T_{b}}$. This suggests that the CBG's useful memory is governed by its experimentally measured distributed delay. Hence, over the appropriate range of symbol periods, $MC\propto \frac{1}{T_{b}}$

\begin{table}[htbp]
\centering
\caption{symbol period sensitivity analysis table}
\label{tab:symbol_period_analysis}
\begin{tabular}{p{1cm} p{1cm}}
\toprule
\textbf{$T_{b}~\mathrm{ps}$} & \textbf{$MC$} \\
\midrule
10 & 23.9508 \\
20 & 14.0010 \\
30 & 9.9980 \\
40 & 8.0007 \\
50 & 6.0014 \\
60 & 5.0011 \\
70 & 4.0007 \\
80 & 4.0002 \\
90 & 3.0008 \\
100 & 3.0008 \\
\bottomrule
\end{tabular}
\end{table}

Particularly, for $T_{b}=50~\mathrm{ps}$, Table~\ref{tab:symbol_period_analysis} gives $MC=6.0014$. The measured CBG delay range is approximately $50-300~\mathrm{ps}$. therefore, the relevant symbol delays are approximately $50, 100, 150, 200, 250, 300~\mathrm{ps}$, giving approximately $6$ accessible symbol-delay intervals. This is consistent with the memory capacity results from the simulation in Figure~\ref{fig:mem_capacity_Jaeger} which also gives $MC=6$ symbol delays.

\section{Discussion}
The results presented in this work provide an initial experimental-data-driven indication that a CBG can provide several of the physical characteristics required for passive photonic reservoir computing. 
A central characteristic of the proposed approach is the distributed delay provided intrinsically by the chirped grating. Because different spectral components experience different reflection positions and corresponding propagation delays, the experimentally measured CBG response can be represented by a set of delay-weight pairs $\{h_i,\tau_i\}$.
This distributed delay structure provides a collection of temporally displaced representations of the same input without requiring the explicit construction of a large network of electronic or optical delay elements. The temporal-mixing results further show that these delayed contributions combine to produce a transformed temporal waveform. Thus, the CBG provides a physically compact mechanism for generating multiple delayed representations of an input signal.
The observed memory behaviour is another important property for reservoir operation. More importantly, the Jaeger-style memory-capacity analysis provides a quantitative measure of the ability of the experimentally parameterized CBG reservoir to reconstruct previous input symbols using a trained linear readout. The falling out of memory contribution as the requested delay extends beyond the approximately measured CBG delay window indicates that the memory is finite rather than indefinitely persistent. 
An important advantage of this approach is that the physical reservoir is provided by the CBG itself. In many reservoir computing implementations, the reservoir must be explicitly constructed from a large number of interconnected nodes, delay elements, couplers, nonlinear components, or other physical resources. In contrast, the CBG already possesses a distributed optical response resulting from its chirped structure. Consequently, the reservoir states can be obtained from the experimentally characterized $\{h_i,\tau_i\}$ response rather than by constructing a separate network of artificial reservoir nodes. This could potentially simplify the physical implementation and reduce the complexity associated with realizing a large reservoir.
The passive nature of the CBG also provides a potentially important energy-efficiency advantage. The grating itself does not require active electrical amplification or powered nonlinear processing to generate the distributed delayed optical response. Its temporal transformation arises from passive propagation, reflection, and interference. Therefore, compared with architectures that require many actively driven reservoir elements, a passive CBG-based reservoir could potentially reduce the energy associated with reservoir-state generation.
Overall, the results suggest that a CBG can serve as a compact passive physical substrate for reservoir computing, rather than merely as a conventional dispersive optical component. The experimentally characterized device already provides distributed temporal delays, temporal mixing, and fading memory, while the Jaeger-style memory-capacity analysis demonstrates that these physical states can be exploited by a trained readout to recover information from previous input symbols. The principal contribution of the present work is therefore not the demonstration of a complete reservoir computer, but the identification and experimental-data-driven characterization of the physical reservoir properties present in a CBG.
These results establish a foundation for future investigation of nonlinear CBG reservoirs, in which the experimentally demonstrated distributed-delay mechanism could be combined with optical nonlinearity, feedback, or other device-level physical mechanisms. Such an architecture could potentially move the CBG from a passive temporal information-preserving element toward a more computationally capable photonic reservoir, with applications in low-latency signal processing, neuromorphic photonic computing, etc. Further work will therefore focus on experimentally realizing the nonlinear regime, quantifying nonlinear separation and computational capacity, and ultimately evaluating the complete system, including optical source and readout overhead in terms of latency, energy consumption, scalability, and task-level performance.

\section{Conclusion}
This work has demonstrated, using experimentally characterised CBG data, that a CBG provides several physical characteristics relevant to passive photonic reservoir computing, including distributed delay, temporal mixing, and finite linear memory. The measured delay range of approximately $50-300~\mathrm{ps}$ was found to determine the temporal scale and number of reconstructable past input symbols. These results establish the CBG as a compact passive physical substrate for reservoir-like temporal processing, while nonlinear separation and complete task-level reservoir computing remain subjects for future investigation.

\section{Data Availability Statement}
The data supporting the findings of this study are available from the author upon reasonable request, subject to reasonable research and intellectual property considerations.

\section*{Acknowledgment}
I sincerely thank my parents for their unwavering prayers, encouragement, and support throughout this research journey. I also gratefully acknowledge Dr. Peter David Girouard for providing me with a strong foundation in academic research and for his guidance in shaping my research career. Finally, my heartfelt gratitude goes to my colleagues who have been encouraging me to continue my independent research.

\ifCLASSOPTIONcaptionsoff
  \newpage
\fi

\bibliographystyle{IEEEtran}
\bibliography{references}
\end{document}